\documentclass[conference]{IEEEtran}
\IEEEoverridecommandlockouts

\usepackage{cite}
\usepackage{amsmath,amssymb,amsfonts}
\usepackage{algorithmic}
\usepackage{graphicx}
\usepackage{textcomp}
\usepackage{xcolor}
\usepackage{caption}
\usepackage{colortbl}
\usepackage{booktabs}
\usepackage{hyperref}
\usepackage{url}
\usepackage{wrapfig}
\usepackage{multirow}
\usepackage{marvosym}
\newcommand{\squeezeup}{\vspace{-2.0mm}}
\PassOptionsToPackage{table,xcdraw}{xcolor}
\usepackage[table,xcdraw]{xcolor} 

\def\BibTeX{{\rm B\kern-.05em{\sc i\kern-.025em b}\kern-.08em
    T\kern-.1667em\lower.7ex\hbox{E}\kern-.125emX}}
\newcommand{\Am}[1]{\mathbf{A}_{#1}}
\newcommand{\Bm}[1]{\mathbf{B}_{#1}}
\newcommand{\Cm}[1]{\mathbf{C}_{#1}}

\begin{document}

\title{State-Space Large Audio Language Models 
}
\author{Saurabhchand Bhati$^{1}$, Yuan Gong$^{1}$, Leonid Karlinsky$^{2,3}$, Hilde Kuehne$^{3,4}$, Rogerio Feris$^{2,3}$, James Glass$^{1}$ \\
$^{1}$MIT, USA, $^{2}$IBM Research AI, USA, $^{3}$MIT-IBM Watson AI Lab, USA, $^{4}$University of Bonn, Germany \\
\small{sbhati@mit.edu}}

\maketitle

\begin{abstract}
Large Audio Language Models (LALM) combine the audio perception models and the Large Language Models (LLM) and show a remarkable ability to reason about the input audio, infer the meaning, and understand the intent. However, these systems rely on Transformers which scale quadratically with the input sequence lengths which poses computational challenges in deploying these systems in memory and time-constrained scenarios. Recently, the state-space models (SSMs) have emerged as an alternative to transformer networks. 

While there have been successful attempts to replace transformer-based audio perception models with state-space ones, state-space-based LALMs remain unexplored. First, we begin by replacing the transformer-based audio perception module and then replace the transformer-based LLM and propose the first state-space-based LALM. Experimental results demonstrate that space-based LALM despite having a significantly lower number of parameters performs competitively with transformer-based LALMs on close-ended tasks on a variety of datasets. 
\end{abstract}

\begin{IEEEkeywords}
State-space models, Large audio language models, Audio reasoning 
\end{IEEEkeywords}

\section{Introduction}

Large Language Models (LLMs)~\cite{devlin2018bert,raffel2020exploring,brown2020language,ouyang2022training,zhang2022opt,zhao2023survey} built using the powerful transformer architecture and trained on web-scale text data have made significant progress in Natural Language Processing. LLMs can read, write understand, and reason about the world through text. ChatGPT and similar systems are increasingly becoming commonplace. 

While text-based systems have made significant strides, in our day-to-day lives we are surrounded by complex audio signals. Understanding audio is a crucial part of developing the next generation of intelligent systems that can interact with the world. Motivated by this, there have been attempts to supplement the LLMs with the ability to understand audio and speech~\cite{deshmukh2023pengi,tang2023salmonn,gong2023listen,gong2023joint,huang2024audiogpt,ghosh2024gama}.

Despite the success of Transformers~\cite{vaswani2017attention}, they suffer from the quadratic time and memory complexity which is a bottleneck for long audio, and speech signals and use on devices with low computational resources. While substantial effort has gone into reducing the computational requirements of the transformer architectures~\cite{wang2020linformer,liu2021swin}, there is a need to explore alternative models for use cases with memory and time constraints. 

State-space models (SSMs) have emerged as an alternative to transformer-based models~\cite{gu2021efficiently,gu2023mamba,liu2024vmamba,bhati2024dass}. SSMs have linear complexity in token length and perform on par with transformers. SSMs demonstrate faster inference time and lower memory requirements than the transformer-based models~\cite{mehta2022long,zhu2024vision,liu2024vmamba}. 
SSMs have demonstrated their performance on text~\cite{gu2023mamba} and image~\cite{zhu2024vision,liu2024vmamba} and are increasingly becoming common for modeling speech and audio signals~\cite{hamza2024audio,lin2024audio,shams2024ssamba,bhati2024dass}. Recently, SSMs coupled with knowledge distillation have been shown to outperform transformer-based teachers and achieve state-of-the-art performance on the audio perception task~\cite{bhati2024dass}.

While there have been works exploring state-space-based audio perception systems, there have been no attempts to explore state-space-based LALMs. In this paper, we systematically explore the impact of using state-space-based systems in developing LALMs. First, we replace the audio perception module from a transformer-based system, AST, to a state-space-based system, DASS, and keep the transformer-based LLM. Then, we also replace the LLMs with state-space LLMs and propose the (to the best of our knowledge) first state-space LALM. We evaluate the model using a mix of datasets for classification, and caption retrieval tasks and show that the state-space LALM performs competitively with transformer-based LALMs. 

\section{Related Work}
LLMs trained on wed-scale data with the next token prediction task showed impressive reasoning and understanding of world knowledge.  These models learn general-purpose representations that can be aligned to desired responses via instruction tuning~\cite{zhang2023instruction}. Large Audio Language Models extend the capabilities of these models beyond text to include general audio and speech. 

Pengi~\cite{deshmukh2023pengi} uses hierarchical transformer HTSAT~\cite{chen2022hts} as the audio encoder, CLIP text encoder as the text encoder, and GPT2~\cite{radford2019language} as the language model. AudioGPT~\cite{huang2024audiogpt} augments ChatGPT's ability to handle complex audio and speech tasks. AudioGPT analyzes the user prompt and assigns a model based on the prompt for example for speech recognition task, whisper~\cite{radford2023robust} is used whereas for speech enhancement ConvTasNet~\cite{luo2019conv} is used. SALMONN~\cite{tang2023salmonn} uses dual encoders: a whisper model to extract speech and information about background noises and a BEATs encoder~\cite{chen2022beats} to extract high-level non-speech audio semantics information and LLaMA as the language-model. 

LTU~\cite{gong2023listen} uses AST~\cite{gong2021ast} as the audio encoder and the LLaMA as the language model. LTU outperforms the existing LALMs on the close-ended task and shows free-form open-ended question-answering capabilities. Our approaches build upon LTU and reduce the computational cost of the model by using state-space models while retaining the performance of these models.  GAMA~\cite{ghosh2024gama} is a concurrent approach that builds upon LTU and combines information extracted from various layers from AST via an Audio Q-Former. GAMA contains significantly more trainable parameters ($ \sim$300M) compared to LTU ($\sim$100M) and our proposed models ($ \sim$40M, $ \sim$60M). GAMA also proposed and used an improved instruction tuning dataset to improve complex reasoning on the input audio. This work explores alternatives for Transformers backbone and builds computationally efficient audio-language generation systems.  

\section{State-space Large Audio Language Models}

\subsection{State-Space Models}

Structured state space sequence models (S4)~\cite{gu2021efficiently} are inspired by classical state-space models such as Kalman filters and Hidden Markov Models. 
The state-space models map a 1-D sequence $ x(t) \in \mathbb{R} \rightarrow y(t) \in \mathbb{R}$ through a hidden state $h(t) \in \mathbb{R}^{N}$ via linear ordinary differential equations as follows:
\begin{align}
    & \mathbf{h'(t)} = \Am{}\mathbf{h(t)} + \Bm{}x(t), \\
    & y(t) = \Cm{}\mathbf{h(t)}   
\end{align}
where $ \mathbf{A} \in \mathbb{R}^{N\times N}$, $( \mathbf{B} \in \mathbb{R}^{N\times 1}, \mathbf{C} \in \mathbb{R}^{1\times N} )$ are called the evolution and projection parameters respectively.

A discretization step transforms the continuous parameters, $\mathbf{A},\mathbf{B} $ to discrete parameters $\bar {\Am{}},\bar {\Bm{}}$. A commonly used method for discretization, zero-order hold uses a timescale parameter $\Delta$ to discretize as follows: 
\begin{align}
    &\bar {\Am{}} = \exp(\Delta \Am{}), \\
    &\bar {\Bm{}} = (\Delta \Bm{})^{-1}(\exp(\Delta \Am{}) - \mathbf{I})\Delta \Bm{}
\end{align}

After the discretization step, the state-space equations can written as:
\begin{align}
    & h_{t} = \overline{\Am{}}h_{t-1} + \overline{\Bm{}} x_{t} \\
    & y_{t} = \Cm{} h_{t}
\end{align}

The current view of state-space models is analogous to RNNs where the output of the current time step only depends on the previous hidden state and current input. Since the state-space parameters are not time or input-dependent, SSM can also be viewed as a convolution $ y_{t} = (x_{0},x_{1},...,x_{t}) * (\Cm{}\overline{\Bm{}},\Cm{}\overline{\Am{}\Bm{}},...,\Cm{}\overline{\Am{}}^{M-1}\overline{\Bm{}}) = \mathbf{x} * \overline{\mathbf{K}} $ where $\overline{\mathbf{K}}$ is the called the global convolutional kernel. 

SSMs can either be viewed as CNNs or RNNs depending on the task.
During training, the convolutional view is used to enable parallel training. During inference, the recurrent view allows faster inference and unbounded context. However, the linear time-invariant nature of these models limits their performance on content-based reasoning tasks. 

Gu et al.~\cite{gu2023mamba} proposed a parametrization method to make the state-space parameters input-dependent. However, this breaks the convolutional view and poses a challenge for efficient computation. To address this, a hardware-aware parallel algorithm is used to efficiently compute the output.

\begin{figure}
    \centering
    \includegraphics[width=0.9\linewidth]{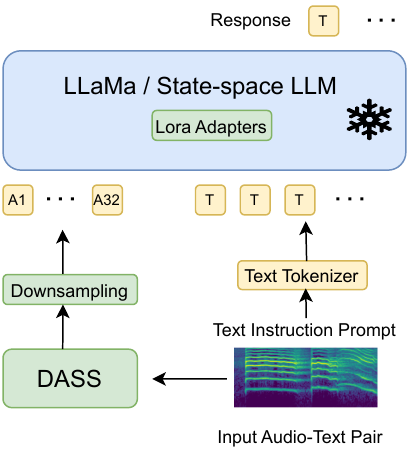}
    \caption{Overview of the proposed State-space Large Audio Language Model. The components in green i.e. DASS, downsampling modules, and the Lora adapters are trainable where the LLM is frozen.}
    \label{fig:ssLALM}
\end{figure}

\begin{table*}[t!]
\caption{Performance comparison on closed-ended audio tasks. ZS: Zero-shot evaluation, the entire dataset is not seen in the training; ZS-: Weak zero-shot evaluation, the dataset is not used in training, but it is sourced from the same project as part of the training data. $^\dagger$ Mean average precision (mAP) underestimates the performance of LALM as they do not predict the likelihood of non-prominent sound classes. $^\ddagger$ Use a higher sampling rate of 44.1KHz than 16KHz used in other approaches.}

\label{tab:close_res}
\captionsetup{skip=1em}
\centering
\resizebox{\textwidth}{!}{
\begin{tabular}{@{}lllllllllllll@{}}
\toprule
\multicolumn{1}{c}{Model}& \begin{tabular}[c]{@{}c@{}}ESC50\textsuperscript{ZS-} \\(Acc)\end{tabular} & \begin{tabular}[c]{@{}c@{}}DCASE\textsuperscript{ZS-} \\ (Mi-F1)\end{tabular} & \begin{tabular}[c]{@{}c@{}}VS\textsuperscript{ZS} \\ (Acc)\end{tabular} & \begin{tabular}[c]{@{}c@{}}TUT\textsuperscript{ZS} \\ (Acc)\end{tabular} & \begin{tabular}[c]{@{}c@{}}BJO\textsuperscript{ZS} \\ (Acc)\end{tabular} & \multicolumn{1}{l}{\begin{tabular}[c]{@{}l@{}}VGG \\ (Acc)\end{tabular}} & { \begin{tabular}[c]{@{}c@{}}FSD$^\dagger$ \\ (mAP)\end{tabular}} & { \begin{tabular}[c]{@{}c@{}}AudioSet$^\dagger$ \\ (mAP)\end{tabular}} & \begin{tabular}[c]{@{}c@{}}Classif.\\ Avg.\end{tabular} & \begin{tabular}[c]{@{}c@{}}AudioCaps \\(SPICE)\end{tabular} & \begin{tabular}[c]{@{}c@{}}Clotho \\(SPICE)\end{tabular} & \begin{tabular}[c]{@{}c@{}}Cap.\\ Avg.\end{tabular} \\ \midrule
\multicolumn{13}{l}{{\color[HTML]{656565} \textit{Best specialized models trained supervisedly on each dataset. Not generalizable to unseen label sets and tasks.}}}           \\
{\color[HTML]{656565} Best Supervised \& Specialized} & {\color[HTML]{656565} 97.0} & {\color[HTML]{656565} 64.6 } & {\color[HTML]{656565} 98.0} & {\color[HTML]{656565} 74.6} & {\color[HTML]{656565} 97.5} & {\color[HTML]{656565} 59.5} & {\color[HTML]{656565} 56.2 } & {\color[HTML]{656565} 47.3 } & {\color[HTML]{656565} 74.3 } & {\color[HTML]{656565} 17.7 } & {\color[HTML]{656565} 13.5 } & {\color[HTML]{656565} 15.6}
\\ \midrule
\multicolumn{13}{l}{\textit{CLIP-like audio-text model. Generalizable to unseen labels, but a pre-defined label set is required for inference.}}              \\
AudioCLIP~\cite{guzhov2022audioclip}                & 69.4                       & -    & -& -& -& -                    & {-}& {25.9}& -    & -        & -     & -\\
Wu et. al~\cite{wu2023large} $^\ddagger$                    & 89.1                       & - & -                    & -                     & -                     & -                    & -                     & - & - & -        & -     & -\\
CLAP~\cite{elizalde2023clap} $^\ddagger$                     & 82.6                       & 30.0& 49.5                    & 29.6                     & 47.5                     & -                    & {30.2}                     & {5.8} & 40.7& -        & -     & -\\  \midrule
\multicolumn{13}{l}{\textit{Baseline LALM: directly output label names, no pre-defined label set is required at inference.}}  \\
\rowcolor[HTML]{EFEFEF} 
LTU (7B)~\cite{gong2023listen}                      & 83.1        & 45.9 & 55.6                  & 32.5                   & \textbf{69.9}                    & 50.3               & {46.3}                     & {18.7}& 50.3& 17.0   & 11.9& 14.5                      \\ 

SALMONN (7B)~\cite{tang2023salmonn} & 16.4\textsuperscript{ZS-} & 18.0\textsuperscript{ZS-} & 16.9\textsuperscript{ZS-} & 7.8\textsuperscript{ZS-} & 25.0\textsuperscript{ZS-} &23.3\textsuperscript{ZS-} & 22.1\textsuperscript{ZS-} & 13.4\textsuperscript{ZS-} & 17.9 &8.3 &7.6 & 8.0\\

Pengi~\cite{deshmukh2023pengi} & 80.8\textsuperscript{ZS-} & 29.6\textsuperscript{ZS-} & 46.4\textsuperscript{ZS-} & 18.4\textsuperscript{ZS-} & 47.3\textsuperscript{ZS-} &16.6\textsuperscript{ZS-} & 35.8 & 11.5 & 35.8& 12.7 &7.0 & 9.9\\

AudioGPT~\cite{huang2024audiogpt} & 41.3 & 20.9 & 35.8 & 14.9 & 21.6 & 5.6 & 18.8 & 12.7 & 21.5 & 6.9 & 6.2 & 6.6\\

\midrule
\multicolumn{13}{l}{\textit{state-space audio encoder (DASS) + LLaMA}}  \\
Small Hybrid-LALM (7B)& \textbf{87.4} & {47.9} & 58.2 & 28.0 & 67.0 & 48.6 & 46.4 & 18.4 & 50.2 & 17.0 & \textbf{12.6} & 14.8\\
Medium Hybrid-LALM (7B)& 85.6 & \textbf{49.5} & 59.4 & 30.6 & 59.3 &  49.4 & 46.1 & 18.3 & 49.8 & 17.6 & 12.4 & 15.0\\

\midrule
\multicolumn{13}{l}{\textit{State-space audio encoder (DASS) + state-space LLM}}  \\

Small ssLALM (3B) & 84.3 & 46.4 & 55.8 & 34.1 & 61.9 & \textbf{51.2} & \textbf{47.8} & 18.6 & 50.0 & \textbf{18.0} & {12.1} & \textbf{15.1}\\

Medium ssLALM (3B) & 86.8 & {47.9} & \textbf{61.2} & \textbf{35.9} & 61.0 & 51.0 & 47.7 & \textbf{19.4} & \textbf{51.4} & {17.7} & 11.7 & 14.7\\

\bottomrule
\end{tabular}
}
\squeezeup
\end{table*}

\subsection{DASS: Distilled Audio State-Space Model}

DASS~\cite{bhati2024dass} was one of the first attempts to use a pure state-space-based model to classify audio signals. DASS uses AST~\cite{gong2021ast}, a transformer-based teacher model, to guide and train a state-space audio classifier. DASS combines the best of both worlds: it outperforms the transformer-based models and retains the computational advantages of state-space models. 

The DASS model can be divided into four groups and each group consists of a state-space block and the first three groups also contain a downsampling layer. Each group progressively reduces the sequence lengths and increases the feature size. A pooling method generates the final embedding that summarizes the input spectrogram. DASS shows remarkable duration scalability: even a model trained with ten-second utterances can infer information from hour-long audio. 

In this work, we use DASS pretrained on AudioSet-2M~\cite{gemmeke2017audio} dataset with the classification layer removed as the audio features extractor for the input audio. It takes a spectrogram of size 1024*128 as input and generates a feature map of size 32*4*768 as output. To further reduce the spatial dimension, we use a two-dimensional convolution with a kernel size of 3 and stride 2 and then use a linear layer to map the features from 768 dimensions to the input size of the language model i.e. 4096 for the LLaMa and 2560 for the state-space based LLM.

\subsection{LLM}

We use the following LLMs in this work:

\textbf{LLaMA}: We follow LTU and use Vicuna instruction tuned LLaMA-7B LLM~\cite{touvron2023llama}. LLaMa is pretrained on large amounts of natural language and code corpora in a self-supervised manner. Vicuna~\cite{chiang2023vicuna} is trained on instruction-following language prompts generated by GPT models which improves the models' performance on reasoning and generation tasks. 

\textbf{State-space LLM}: State-space LLM-2.8B~\cite{gu2023mamba} trained on the Pile dataset. The state-space LLM outperforms similar-size transformer-based models such as GPT-Neo 2.7B. 

\textbf{Low-rank Adapters}: Instead of finetuning all the weights of LLMs on our task, we use Low-rank (LoRA)~\cite{hu2021lora} adapters to finetune the LLMs. LoRA adds a small set of learnable weights on top of the pre-trained weights from the LLM. The learnable weights can be decomposed into a product of two low-rank matrices. This allows us to modify the large parameter matrices of the LLMs without adding a lot of learnable parameters. The final LLM parameters are the addition of frozen parameters and the low-rank learnable matrices. 

For LLaMa, we add LoRA adapters (rank=8 and $\alpha$=16) to the key and query projection layers in all the self-attention layers of LLaMa models. This step introduces 4.2M learnable parameters. For state-space LLM, we add LoRA adapters (rank=8 and $\alpha$=16) to the input projection layers of the state-space block. This step adds 6.5M learnable parameters.

\subsection{Training Objective}
We train our models using the next token prediction task conditioned on the input audio and past tokens. We maximize the following probability, $P(x_t | x_{1:t-1}, A)$, by using cross-entropy loss for all the text tokens $1 < t \leq T$ in the input text tokens and reference audio. For generation, we use the following settings: Temperature=0.1, Top K=500, and Top P=0.95 with a repetition penalty of 1.1.

\subsection{Experiments}

We train our models on \texttt{OpenAQA} dataset~\cite{gong2023listen}. This dataset contains tuples of \texttt{Audio, question, answer} where the models take \texttt{Audio} and \texttt{question} as input and generate \texttt{answer} as the output.

We follow the same training pipeline from LTU for training the models proposed in this paper. We use 4× RTX A6000 GPUs for training the models. The hybrid LALM is trained for about 3 days. For the state-space LALM, we can increase the batch size from 4 to 16. We use gradient accumulation to ensure the effective batch size is the same i.e. 256. This speeds up the training and allows us to train the model in less than two days. However, the reason for the training speed up is unclear: it could be simply due to the state-space LLM being smaller or the state-space model being computationally efficient. In the future, when larger state-space LLMs become publicly available, we plan to increase the model size and compare the computational efficiencies of the state-space and Transformer-based LALMs. 

The small and medium hybrid-LALM contains 42M and 61M trainable parameters out of 6.8B parameters respectively. The small and medium ssLALM contains 43M and 62M trainable parameters out of a total of 2.8B parameters respectively.

\begin{table*}[t!]
    \centering
    \begin{tabular}{p{17.5cm}}\toprule 
    {\color[HTML]{3531FF} Question: Based on the combination of sounds in the audio clip, what kind of atmosphere or environment does it evoke? }{\color[HTML]{656565} [Scene Understanding]}\\
    {\color[HTML]{CB0000}Reference: The combination of sounds suggests a rural or natural setting with animals and machinery.} \\ \midrule 
    LTU: The combination of sounds in the audio clip evokes a busy, active atmosphere, possibly a quirky environment with a mix of mechanical devices, talking, an animal, and tapping. \\ \midrule
    Small Hybrid-LALM: The audio clip evokes a natural and outdoor atmosphere with animals and people interacting together.  \\
    Medium Hybrid-LALM: The combination of sounds in the audio clip evokes a rural or natural environment with animals present. \\ \midrule 
    Small ssLALM:  The combination of sounds in the audio clip evokes a busy, active atmosphere, possibly a quirky environment with a mix of mechanical devices, talking, an animal, and tapping.  \\
    Medium ssLALM: The combination of sounds creates a lively and bustling atmosphere with people talking while pigs are making noise.   \\ 
    
    \midrule \midrule 

    {\color[HTML]{3531FF} Question: Based on the audio clip's acoustic features, how can we differentiate between Liquid and Gurgling, bubbling sounds?} {\color[HTML]{656565} [Detailed Perception Question]}\\
    {\color[HTML]{CB0000}Reference: Liquid sounds are characterized by Smooth, flowing, and consonant features, while Gurgling, bubbling sounds have a Low-pitched, rhythmic, wet popping sensation.} \\ \midrule 
    LTU: Liquid sounds are characterized by Smooth, flowing, and consonant features, while Gurgling, bubbling sounds have a Low-pitched, rhythmic, wet popping sensation. \\ \midrule
    Small Hybrid-LALM: Liquid sound is smooth and flowing while Gurgling, bubbling sound is more irregular and intermittent.  \\
    Medium Hybrid-LALM: Liquid sound is characterized by a smooth and continuous flow of water, while gurgling and bubbling sounds are caused by air bubbles in the liquid that create turbulence and noise. \\ \midrule 
    Small ssLALM:  Liquid sound is smooth with a continuous flow of water while gurgling or bubbling noise has intermittent bursts.  \\
    Medium ssLALM: Liquid sound is smooth and flowing while gurgling has a distinct bubbly quality.   \\

    \bottomrule
    \end{tabular}
    \caption{Open-ended AQA samples for the LALMs proposed in the paper.}
    \label{tab:my_label}
\end{table*}

To reason and understand the audio, the LALM must be able to first recognize the input audio. One important step in bench-marking LALMs is the performance comparison on close-ended tasks where the output labels are predefined. We follow the evaluation pipeline from LTU~\cite{gong2023listen} and compare our models with existing LALMs on 8 audio classification benchmarks and 2 audio captioning benchmarks. 

\textbf{Audio Classification}: LALMs do not directly predict the class index but instead output the audio label names or descriptions. To compute the performance of these models, we first encode the LALM output and the evaluation label using a text encoder and then we compute a cosine similarity between the LALM output and the label. For single-label classification tasks, we use the label with the highest similarity score and compute accuracy or F1-score and for multi-label classification tasks, we use the cosine similarity as the prediction score and compute the mAP. We used the prompt ``write an audio caption describing the sound" for the classification tasks. 

\textbf{Audio Captioning}: For the caption generation tasks, we use the prompt ``write an audio caption describing the sound” and take the LALM output as the prediction. We AudioCaps and Clotho datasets and use SPICE as the evaluation metric.

As seen in Table 1, our proposed model outperforms the other existing models such as SALMONN, Pengi, and AudioGPT. For SALMONN~\cite{tang2023salmonn}, Pengi~\cite{deshmukh2023pengi}, AudioGPT~\cite{huang2024audiogpt}, we use the results reported in the GAMA paper. 
For both the audio classification and audio captioning task our proposed models outperform the existing models in most of the datasets and overall average performance. For the audio captioning task, our models perform similarly to the best-supervised systems.

Our proposed models match the performance of LTU~\cite{gong2023listen} which our approaches build upon. The audio encoder i.e. AST used in LTU is first pretrained on audio-visual data and then finetuned on the AudioSet-2M dataset. In contrast, the DASS model is trained on the audio data from AudioSet-2M.
DASS trained on only AudioSet-2M outperforms and shows more robustness and duration-scalability than AST trained on the same dataset~\cite{bhati2024dass}. 

The state-space LLM used in ssLALM is much smaller and is not instruction-trained unlike LLaMA used in hybrid LALMs or transformer LALMs such as LTU. Despite being smaller and having the audio encoder and LLM trained on less data, the ssLALMs perform competitively with the best transformer-based LALMs. 

Although all the LALMs, including the ssLALMs, perform poorly on multi-label classification task on AudioSet. We believe it is because LALMs mainly predict the prominent class and underestimate the likelihood of non-prominent sound classes in the input audio which results in low mAP scores. We also show some open-ended question-answering abilities of the models in Table 2.  All the models can infer the information from audio and generate reasonable answers. 

\section{Conclusions and Future Work}
In this paper, we propose the first state-space large audio language model. We systematically replace the audio-perception and LLM components in the LALMs and analyze the performance. Our experiments show that ssLALMs perform competitively with the transformer-based LALMs despite using a significantly lower number of parameters. 

In the future, we would like to scale up the data used for training the model and more complex reasoning datasets such as \texttt{CompA-R}~\cite{ghosh2024gama}. We would also like to build the larger ssLALMs with large state-space LLMs. We would also like to explore state-space attention hybrid models such as Jamba as the language models as they combine the best of the state-space and transformers.

\bibliographystyle{IEEEbib}
\bibliography{ref}

\end{document}